\documentclass{article}

\usepackage{a4wide}
\usepackage[centertags]{amstex}
\usepackage{epsfig}             

\makeatletter
\renewcommand\section{\@startsection{section}{1}{\z@}%
                                   {-3.5ex \@plus -1ex \@minus -.2ex}%
                                   {2.3ex \@plus.2ex}%
                                   {\reset@font\Large\scshape}}
\renewcommand\subsection{\@startsection{subsection}{2}{\z@}%
                                     {-3.25ex\@plus -1ex \@minus -.2ex}%
                                     {1.5ex \@plus .2ex}%
                                     {\reset@font\large\slshape}}

\makeatother

\def\ap#1#2#3   {{Ann.\ Phys.\ (NY)}\ #1 (#2) #3}
\def\ib#1#2#3   {ibid., #1 (#2) #3}
\def\np#1#2#3   {{Nucl.\ Phys.}\ #1 (#2) #3}
\def\pl#1#2#3   {{Phys.\ Lett.}\ #1 (#2) #3}
\def\prep#1#2#3 {{Phys.\ Rep.}\ #1 (#2) #3}
\def\prev#1#2#3 {{Phys.\ Rev.}\ #1 (#2) #3}
\def\prl#1#2#3  {{Phys.\ Rev.\ Lett.}\ #1 (#2) #3}
\def\zp#1#2#3  {{Z.\ Phys.}\ #1 (#2) #3}

\newcommand\thi{\ensuremath{\theta_1}}
\newcommand\thii{\ensuremath{\theta_2}}

\newcommand\Sphi{\ensuremath{\sin\frac{\phi}{2}}}
\newcommand\SPhi{\ensuremath{\sin\frac{\Phi}{2}}}

\newcommand\Cphi{\ensuremath{\cos\frac{\phi}{2}}}

\newcommand\half{\ensuremath{\frac{1}{2}}}
\newcommand\Tr{\operatorname{Tr}}

\begin{document}

\thispagestyle{empty}

\begin{center}
July 1998 \hfill \large IFUP-TH 27/98 \\
\hfill hep-lat/9807019
\end{center}

\vspace{1cm plus 1fil}

\begin{center}

\textbf{\huge Wilson loop distributions, higher representations and
  centre dominance in SU(2)}

\vspace{1cm plus 1fil}

{\large
{\bfseries P.W. Stephenson \\}
           Dipartimento di Fisica dell'Universit\`a and INFN, \\
           I-56100 Pisa, Italy \\
           Email: \texttt{pws@@ibmth.df.unipi.it}}

\vspace{1cm plus 1fil}

\textbf{\Large Abstract}

\end{center}

\noindent To help understand the centre dominance picture of
confinement, we look at Wilson loop distributions in pure SU(2)
lattice gauge theory. A strong coupling approximation for the
distribution is developed to use for comparisons.  We perform a
Fourier expansion of the distribution: centre dominance here
corresponds to suppression of odd terms beyond the first.  The Fourier
terms correspond to SU(2) representations; hence Casimir scaling
behaviour leads to centre dominance.  We examine the positive
plaquette model, where only thick vortices are present.  We show that
a simple picture of random, non-interacting centre vortices gives a
string tension about 3/4 of the measured value.  Finally, we attempt
to limit confusion about the adjoint representation.

\vspace{1cm}



\newpage

\section{Introduction}

Recently the idea, originally proposed some time
ago~\cite{Co79,Ho79,MaPe79}, that confinement in gauge theories can be
considered as an effect related to the centre Z($N$) of the gauge
group SU($N$) has been undergoing a
revival~\cite{DDFa97a,DDFa97b,DDFa98,KoTo97}.  Not the least of this
has been the demonstration~\cite{KoTo97} that if one replaces the
values of Wilson loops with their signs alone (i.e.\ the centre Z(2))
the heavy quark potential in SU(2), which is now very accurately
known~\cite{BaSc95}, can be essentially \textit{completely}
reproduced.  This is a spectacular --- and gauge invariant ---
demonstration that something is right about this hypothesis; it is,
however, far from a demonstration that centre vortices are responsible
for confinement.

The key part of the argument for Z(2) to be an ingredient in
confinement is that there are `thick' centre vortices, associated with
the quotient group $\mathrm{SU}(2)/\mathrm{Z}(2) \cong
\mathrm{SO}(3)$, which pierce Wilson loops; their physical effect
depends on how many vortices, modulo 2, pierce a given area.  There
are also `thin' vortices associated with Z(2), which appear as chains
of negative plaquettes; to create these requires an action
proportional to the number of flipped plaquettes, so these will not
survive in the continuum limit.  By contrast, the thick vortices show
up only in larger Wilson loops, which may be negative while still
surrounding only positive plaquettes, and do survive in the continuum
limit; they are topological in nature, related to the fact that
SU(2)/Z(2) is not simply connected and indeed has a Z(2) homotopy.
(For simplicity, we have here ignored hybrid vortices, which combine
the two effects.)  We shall return to the distinction between Z(2) and
SU(2)/Z(2) effects towards the end of the article because, as
emphasised in ref.~\cite{KoTo97} where this mechanism is described in
more detail, it is important and has caused much confusion.

We have specialised to SU($N$) for $N=2$ since the arguments are
expected to extend to higher $N$ via the centre Z($N$), though of
course this needs to be checked explicitly.  We are also ignoring the
effects of fermions in the vacuum.

\begin{figure}
  \begin{center}
    \psfig{file=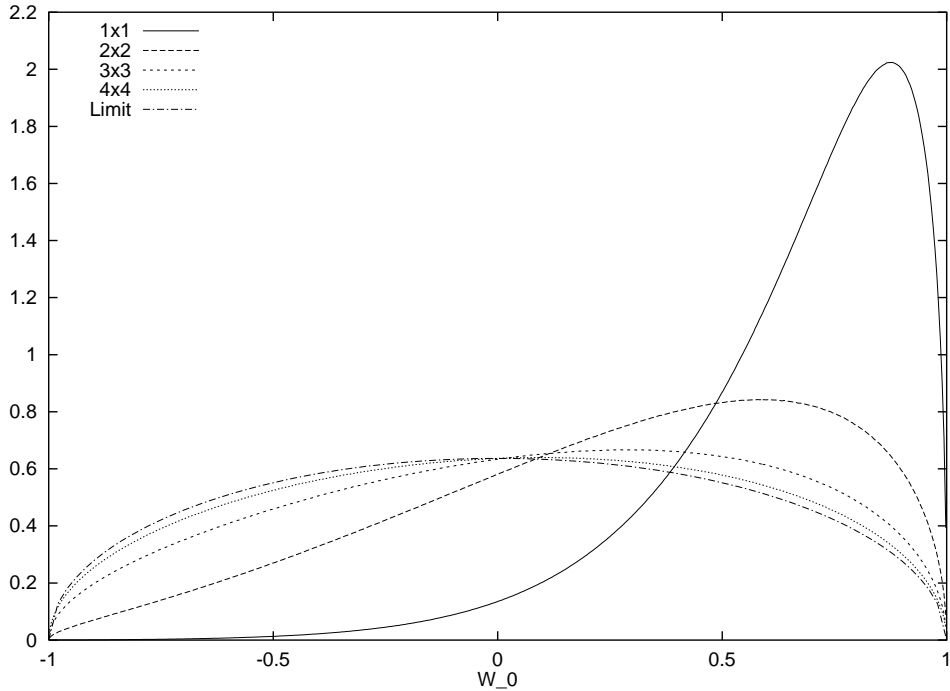,width=3.7in,angle=270}
  \end{center}
  \caption{Monte Carlo distribution of Wilson loops on a $12^4$
    lattice at $\beta=2.5$, showing also the limit of the distribution
    for large loops.  The area under each curve is normalised to
    unity.  On this scale errors are negligible.}
  \label{fig:wl_dist}
\end{figure}

Because Z(2) commutes with all elements of the gauge group, the sign
of any Wilson loop in the fundamental representation of SU(2) must be
determined by the combined effect of thin and thick vortices: the
total of these objects passing through the loop determines whether the
loop is positive (even number of vortices) or negative (odd number).
In an attempt to see how this appears, we can look at the distribution
of the trace of the Wilson loop, as generated by a Monte Carlo
simulation.  Note we are looking at the distribution of individual,
not average, Wilson loops.  This was first looked at, in the case of
the plaquette, in refs.~\cite{MaPo82,BeMa84}, where it is referred to
as the `spectral density', and the point was made that it contains
information about all representations.  One such set, taken from
10,000 configurations of a $12^4$ lattice at $\beta=2.5$, is shown in
fig.~\ref{fig:wl_dist}.  It is clear that it changes smoothly from the
highly asymmetric shape of the plaquette action to some limiting
distribution for large loops.  It is less clear how the Wilson loop
expectation value $\langle W_0\rangle$, which is the first moment of
the distribution, behaves; nor is it clear where the special role of
the sign of the loop comes from.

This paper attempts to shed some light on these matters.  As the
behaviour of the distribution is largely unfamiliar territory, we
first perform a calculation using a simple approximation: we are not
able to calculate the raw distribution, but we are able to see how it
develops for larger loops.  The expectation value in this
approximation has the leading order strong coupling behaviour.

We shall proceed as follows: first (section 2) we shall give a few
basic home truths about Wilson loops.  Next (section 3), we shall look
at the approximation that all loops are uncorrelated, corresponding to
strong coupling; in section 4 we discuss the centre dominance picture
in the same spirit, where we also explicitly show the connection
between the shape of the distribution, irreducible representations of
SU(2) and the phenomenon of centre dominance.  Then we make some
remarks about centre-projected vortices, the positive plaquette model,
the adjoint representation, and show how a gas of non-interacting Z(2)
vortices gives exact area law confinement.  In the last section we
summarise these results concisely.

Section 3 can safely be missed out by anyone not interested in the
details of the distributions of uncorrelated Wilson loops.  Of the
formalism in that section we shall only use the formula
relating the distributions for loops of areas $A$ and $2A$,
eqn.~(\ref{eq:fmult}), and its extension to arbitrary multiples of
$A$.  The main physical discussion is in section 4.

\section{Wilson loop distributions}

We shall start by assuming the standard Wilson lattice gauge theory in
SU(2).  We parametrise an open Wilson loop (i.e.\ the SU(2)
element representing the loop, where no trace has been taken) as
\begin{equation}
W \equiv W_0 \mathbf{1}_2 + i\mathbf{\sigma}.\mathbf{W},
\qquad W_0^2 + W_1^2 + W_2^2 + W_3^2 = 1.
\end{equation}

We now consider the distribution, the word being used in the sense of
a Monte Carlo calculation: we have a large number of configurations,
and consider the spread of values for all Wilson loops of a given
shape on all lattices.  Of course, we could equally well talk in terms
of contributions to the path integral, where, because of the
importance sampling built into the Monte Carlo simulations, the
exponential factor is included and there is a uniform measure; this is
entirely equivalent.  However, the statistical language fits well here
where we use actual Monte Carlo data for comparison.  The distribution
contains all the gauge invariant information about the loops,
including the expectation values of the loops in all representations.
Indeed, in a quantum theory it is natural to consider this
distribution as the basic quantity.

The first matter of interest is the limiting large loop distribution
apparent in fig.~\ref{fig:wl_dist}.  This can easily be calculated.
Given the short range behaviour of the force, for sufficiently large
loops different parts of the loop are physically unconnected with one
another.  The lack of any overall correlation means the loops are
random: the distribution corresponds to a random walk by $W_i$ over
the 3-sphere of the gauge manifold, so we simply
need to calculate the fraction of the surface available at a given
$W_0$ with the other co-ordinates fixed.  The probability that $W_0$
lies in a certain range is
\begin{equation}
  \rho_l(W_0)\,dW_0 = S_0 ds_0 / S,
\end{equation}
where here, as throughout, we use the symbol $\rho$ for the
distribution and the suffix $l$ indicates the limiting distribution, 
$S_0$ is the 2-surface for a given $W_0$, while between $W_0$ and
$W_0+dW$ a distance $ds_0$ is covered on the surface of the 3-sphere
with the other co-ordinates held constant, and $S$ is the total
surface area of the 3-sphere.  Basic geometry gives
\begin{equation}
  \rho_l(W_0)\,dW_0
  = \frac{4\pi(1-W_0^2)}{2\pi^2}\frac{dW_0}{(1-W_0^2)^{1/2}}
  = \frac{2}{\pi}(1-W_0^2)^{1/2}dW_0.
  \label{eq:limiting}
\end{equation}
This is the lowest curve in fig.~\ref{fig:wl_dist} and clearly fits
the part.  Indeed, loops larger than $4\times4$ are so close to this
that they have been omitted for clarity.
Later, we will see that this curve corresponds to the identity
representation of SU(2); it will then be obvious that convergence
to this limit is inevitable provided that all $j=1/2$ and higher
representation Wilson loops decay monotonically.

The loop distribution in figure~\ref{fig:wl_dist} is smooth and we
would expect to be able to expand it in a Fourier series.  We will
pick the units $W_0\equiv \Sphi$, where $-\pi\le \phi\le\pi$ so that
the terms are orthogonal on the range of values of $W_0$.
However, our expansion is not quite standard; we use our freedom to
expand the even and odd parts separately to write,
\begin{equation}
  \rho_A(W_0) \equiv \rho_A(\Sphi) = 
  \sum_{n=1}^{\infty} (a_n\cos (n-\half)\phi + b_n\sin n\phi)
  \label{eq:fourier}
\end{equation}
as the distribution of $W_0$ at loop area $A$.
The difference from the usual Fourier series is
that this form satisfies the boundary condition that the function
vanishes at $\phi=\pm\pi$.  The basis corresponds to the
eigenfunctions of the self-adjoint linear equation
\begin{equation}
  \frac{d^2\rho}{d\phi^2} + \nu^2\rho = 0
\end{equation}
with those boundary conditions, so that orthogonality (which may
easily be checked) and completeness are guaranteed by Sturm-Liouville
theory.  We will later show that there is a one-to-one correspondence
between these terms and the expectation value in irreducible
representations of the gauge group; this is why we have
picked this form.  This means that any distribution which does not
vanish at the boundary does not have an interpretation in terms of the
underlying group theory.  The large area limit is already visible as the
$a_1$ term.

As a final piece of preparatory formalism, we also calculate the
average Wilson loop in the fundamental representation from the Fourier
series; this is
\begin{multline}
  \langle W_0\rangle = \int_{-1}^1\rho_A(W_0)W_0dW_0
  = \int_{-\pi}^{\pi}\rho_A(W_0)\Sphi\,\Cphi\,d\phi/2 \\
  = (1/4)\int_{-\pi}^{\pi}(b_1\sin\phi)\sin\phi\,d\phi
  = \frac{\pi b_1}{4},
  \label{eq:w0b1}
\end{multline}
since by orthogonality only the first odd term in the expansion survives.
Later we will generalise this formula to all representations.

\section{The uncorrelated loop approximation}
\label{sec:uncor_loop}

We now claim that due to gauge invariance we can always write an
individual Wilson loop (the qualification is important)
chosen from the distribution  as
\begin{equation}
W \equiv W_0^2 + \sqrt{1 - W_0^2}\mathbf{\sigma}.\mathbf{e_r},
\label{eq:wdist}
\end{equation}
where $\mathbf{e_r}$ is a unit vector in a random direction in the
3-space inhabited by $\mathbf{W}$.  This is essentially a
consequence of Elitzur's theorem. To see this, consider any Wilson
loop of a given shape with its opening at a particular point on a
lattice.  If the distribution of $\mathbf{W}$ is not as shown, then
there must be some preferred direction(s) for $\mathbf{W}$.  However,
we can perform a local gauge transformation at the site of the
opening.  It is easy to see that this leaves $W_0$ invariant, while
rotating $\mathbf{W}$ to an arbitrary angle.  As we have complete
freedom to do this, with the single exception that rotations of Wilson
loops with their openings on the same site are correlated, we can move
the preferred direction of $\mathbf{W}$.  (We can avoid the exception
here by simply considering only sets of translated Wilson loops
without rotations, which gives a perfectly valid distribution.)  Thus
there is no preferred direction after all, and the distribution is the
one claimed.

\begin{figure}
  \begin{center}
    \psfig{file=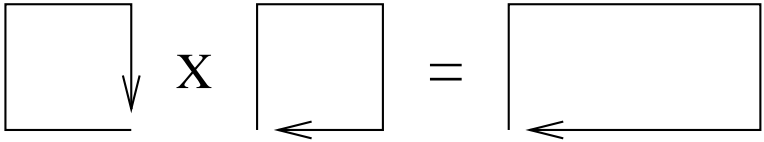,angle=270,width=4in}
  \end{center}
  \caption{Multiplication of adjacent Wilson loops.  We can move the
    opening using a gauge transformation and hence create loops of
    any size in this fashion.}
  \label{fig:wl_mult}
\end{figure}

We now take two Wilson loops, $W^1(A)$ and $W^2(A)$, of the same
shape and area $A$, but chosen in such a way that their opening lies on
the same site, fig.~\ref{fig:wl_mult}.  It will also be convenient to
choose rectangular loops with a side in common traversed in such a way
that this cancels out in the product of the loops.  Thus we produce a
Wilson loop of twice the area.  The $W_0$ element of this loop, by the
standard SU(2) multiplication rule, is:
\begin{equation}
    W_0(2A) = = W_0^1(A)W_0^2(A) - 
    (1-{W_0^1(A)}^2)^{1/2}(1-{W_0^2(A)}^2)^{1/2} \cos\xi,
  \label{eq:wmult}
\end{equation}
where $\cos\xi \equiv \mathbf{e}_r^1.\mathbf{e}_r^2$; $\xi$ is the
angle between the two unit three vectors in the 
directions $\mathbf{W}^1(A)$ and $\mathbf{W}^2(A)$.

The correlations between the diagonal elements $W_0^i(A)$ in
eqn.~(\ref{eq:wmult}) together
with the angle $\xi$ contain all the information about correlations
between adjacent loops; without them we simply have loops from a random
distribution multiplied together.  If we were to go on and consider
larger loops still, we would have to consider the cumulative effect of
these correlations on the larger loops, and the expressions would
rapidly become unmanageable.

Instead, we make the approximation that both the angular and the
diagonal correlations are negligible for loops larger than some area
$A=A_d$.  We shall later see explicitly that this gives the leading
strong coupling value for $\langle W_0\rangle$, although in our case
expanded in terms of smaller Wilson loops rather than the coupling
itself.  That such behaviour is expected is due to the random nature
of the choice of loops from the distribution.

A random distribution for $\cos\xi$ is simply a flat distribution,
which follows from the fact that the usual measure for the polar angle
on a 2-sphere (here the one containing $\mathbf{e}_r^i$) is
$\sin\xi\,d\xi=d\cos\xi$.  We are assuming, however, that the loop
distribution $\rho(W_0)$ itself has not reached the large area limit.
The formula (\ref{eq:wmult}) can now be used to multiply loops of
different sizes, and we can build up larger Wilson loops by repeated
application.  Moving the opening in the Wilson loop simply corresponds
to a gauge transformation, leaving the only remaining variable $W_0$
invariant, so there is no difficulty in constructing Wilson loops of
any area which is a multiple of $A_d$ by this method.

Our procedure is to take the measured Monte Carlo distribution of
$W_0$ on loops of area $A=A_d$ as our input and hence use the
approximation to calculate the distribution of Wilson loops for larger
areas.  In our examples, we shall in fact start from the plaquette,
i.e.\ take $A_d=1$ in lattice units and the measured plaquette
distribution, but this is not forced on us.  This method is obviously
something of a hybrid, since the initial distribution includes all
correlations, but it will allow us to perform explicit calculations
using the loop distribution.  Further, we will be able to predict the
behaviour, starting from the appropriate expectation value for the
plaquette, in every representation of the gauge group.

\subsection{Calculation of the product distribution}

Suppose that the probability distribution of $W_0$ for loops of the
initial size $A_d$ is given by $\rho_{A_d}(W_0)dW_0$, normalised so
that the integral between $W_0=-1$ and 1 is unity.  To produce the
distribution of larger loops we must integrate over the smaller ones,
using the equation~(\ref{eq:wmult}) as a constraint.  It will be
helpful at this point to change to angular variables: take
$W_0(2A_d)\equiv\cos\Theta$, where $0\le\Theta\le\pi$, and likewise
$W_0^i(A_d)\equiv\theta_i$ for $0\le\theta_i\le\pi$.  The new
distribution is given by
\begin{equation}
  \rho_{2A_d}(W_0) dW_0 = \frac{1}{2}\int\int
\rho_{A_d}(\cos\thi)\rho_{A_d}(\cos\thii)\sin\thi d\thi
\sin\thii d\thii\sin\xi d\xi,
\label{eq:intprod}
\end{equation}
where the extra factor of $1/2$ in eqn.~(\ref{eq:intprod}) has appeared
because in covering the full range of $\thi$ and $\thii$ we
cover the $\Theta$ range exactly twice.  This can be seen by
simultaneously exchanging $\thi\leftrightarrow\pi-\thi,
\thii\leftrightarrow\pi-\thii$ in eqn.~(\ref{eq:constraint}), which
leaves $W_0$ invariant.

The range in this notation is taken from eqn.~(\ref{eq:wmult})
which becomes
\begin{equation}
  W_0 \equiv \cos\Theta = \cos\thi\cos\thii - \sin\thi\sin\thii\cos\xi.
  \label{eq:constraint}
\end{equation}

Note that with the correct normalisation of $\rho_{A_d}(\cos\theta)$,
the new distribution will also be correctly normalised, i.e.
$\int\rho_{2A_d}(W_0) dW_0=1$, regardless of the form of $\rho$; this
is nothing more than a basic consistency condition for the integrals
we are performing.

We remove $d\xi$ from eqn.~(\ref{eq:intprod}) using
eqn.~(\ref{eq:constraint}), $\sin\thi\sin\thii\sin\xi d\xi = dW_0$,
giving
\begin{equation}
  \rho_{2A_d}(\cos\Theta) = \frac{1}{2}\int\int_\mathrm{constraint}
  \rho_{A_d}(\cos\thi)\rho_{A_d}(\cos\thii) d\thi d\thii
\end{equation}
with the constraint of eqn.~(\ref{eq:constraint}).
The considerable simplification due to the approximation that
$\cos\xi$ has a random distribution is now evident.

The explicit evaluation of the integral is rather uninteresting.  We
shall simply point out that it is easiest to change variables to the
$\phi$ used in the Fourier series (\ref{eq:fourier}), so that $W_0
\equiv \cos\theta \equiv \Sphi$.  We then insert the Fourier series
for the initial distribution $\rho_{A_d}$ into the integral, and after
some even more tedious algebra best left to computers we obtain,
\begin{equation}
  \rho_{2A_d}(\SPhi) =
  \sum_{n=1}^\infty\frac{(-1)^{n-1}\pi}{4}
    \left(\frac{a_n^2}{n-\half}\cos(n-\half)\Phi
      + \frac{b_n^2}{n}\sin n\Phi \right).
  \label{eq:fmult}
\end{equation}
Both odd and even sets of terms remain separate; as we have already
hinted, this is no coincidence.

The relationship between $b_1$ and the loop expectation value was
given in eqn.~(\ref{eq:w0b1}).  We now read off the coefficient of
$\sin \Phi$ which is $\pi b_1^2/4$, hence the new expectation value is
$(\pi b_1/4)^2$.  Because the terms in the expansion remain separate,
we can immediately extend the result to N multiplications and a loop
size $A\equiv N+1$,
\begin{equation}
\langle W_0(A)\rangle = (\pi b_1/4)^{1+N} = W_0(1)^{A},
 \label{eq:arealaw}
\end{equation}
an area law behaviour with string tension $-\log(W_0(1))$.

Eqn.~(\ref{eq:arealaw}) is equivalent to the lowest order strong
coupling result, even though in this case no mention of the coupling
has been made; indeed, it is renormalised due to the fact that we
start from the physical plaquette rather than the strong coupling
prediction, although the behaviour for larger loops in terms of the
plaquette is the same.

\begin{figure}
  \begin{center}
    \psfig{file=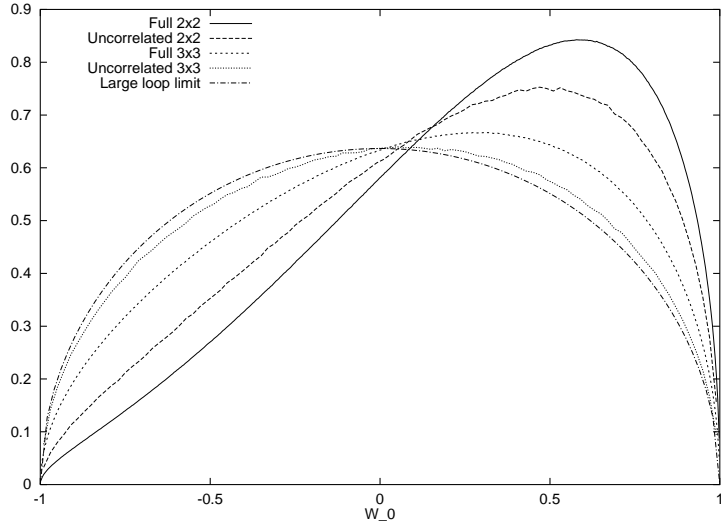,angle=270,width=4in}
  \end{center}
  \caption{Comparison of Wilson loop distributions from Monte Carlo
    and those from the uncorrelated loop approximation (where the
    noise is clearly visible).  The measured plaquette distribution
    has been used as a starting point for the latter.}
  \label{fig:wl_compare}
\end{figure}

In fig.~\ref{fig:wl_compare}, we compare distributions from
uncorrelated loops with the measured values taken from
fig.~\ref{fig:wl_dist}; as already stated, we have started from the
measured plaquette distribution to generate the uncorrelated loop
distributions.  The multiplications in this case were done by Monte
Carlo integration of the rule~(\ref{eq:intprod}), which was rather more
convenient than applying the analytical formulae, thus accounting for
the noise in this distribution.

Not surprisingly, we immediately see the measured shape start to
diverge from the uncorrelated loop shape.  So it is not at all clear
this simple approximation is useful in explaining the physics.
However, later we will see that the terms in eqn.~(\ref{eq:fmult})
correspond to irreducible representations of SU(2) and that the
individual terms separately obey an area law.  Thus the calculation
has the useful effect of demonstrating the leading strong coupling
behaviour in all representations.

\section{Centre dominance}

As explained in the introduction, the centre dominance picture applied
directly to Wilson loops has been shown to be quantitatively
successfully, indeed in very good agreement with the full
values~\cite{KoTo97}.  In this approach any Wilson loop obtained from
Monte Carlo results in SU(2) is approximated by its sign.  It should
be pointed out that in a weak sense it is not surprising that this
reflects aspects of the dynamics: if one divides $\rho(W_0)$ into odd
and even parts, only the odd parts can contribute to $\langle
W_0\rangle$ because that comes from the first moment of the
distribution.  We shall return to this point below.  The result that
the sign \textit{alone} contains the essential dynamics is stronger
and deserves examination.

In terms of the distribution, the centre dominance picture requires
that we count $-1$ for a loop with $-\pi\le\phi\le0$ and $+1$ for
$0\le\phi\le\pi$ with $W_0\equiv\sin(\phi/2)$, i.e.\ a step
distribution.  We use the Fourier expansion of the Wilson loop
distribution in eqn.~(\ref{eq:fourier}) and obtain for the expectation
value in this picture,
\begin{multline}
  \langle W_0^{Z(2)}\rangle = \int_0^\pi\rho\Cphi \frac{d\phi}{2} -
  \int_{-\pi}^0 \rho\Cphi \frac{d\phi}{2} \\
  = \frac{1}{2}\sum_n\left(\int_0^\pi b_n\sin n\phi\Cphi d\phi
  - \int_{-\pi}^0b_n\sin n\phi\Cphi d\phi\right)
  = \sum \frac{nb_n}{n^2-1/4}.
  \label{eq:cdexpn}
\end{multline}

The series here is to be compared with the exact value $\pi b_1/4$.
The important difference is the presence of all the higher $b_n$, with
decreasing coefficients, in the centre dominance formula; without
these the results would differ only by an overall factor of $3\pi/16$,
which vanishes in the ratios required for the heavy quark potential
(see eqn.~(\ref{eq:potdef}), below).  Why the value in
eqn.~(\ref{eq:cdexpn}) gives the correct value can therefore by
clarified by looking at the values of $b_n$ in the real distribution.
Some support for this result is given by the uncorrelated loop
picture, eqn.~(\ref{eq:fmult}), where in doubling the area of a loop
the new coefficients $b_n$ are suppressed by a factor $1/n$, which in
combination with the factor $n/(n^2-1/4)$ in eqn.~(\ref{eq:cdexpn})
shows that higher terms become progressively less important both for
large $A$ and for large $n$.  If the other $b_n$ are already smaller
than $b_1$ for small loops this effect is enhanced because of the
powers of $b_n$ involved.

We have calculated the coefficients $b_n$ by binning Wilson loop data
from Monte Carlo simulations and performing the integral for the
Fourier coefficients,
\begin{equation}
  b_n = \frac{1}{\pi}\int_{-\pi}^\pi\rho\sin n\phi\,d\phi,
\end{equation}
numerically on the data at the end of the run.  Errors have been estimated
by averaging over values of $b_n$ obtained in this way from several
runs.  The results here use 500 bins; we have compared this with the
result from 200 bins in order to check that the
discretisation error due to the finite number of bins is small. This
error increases with the coefficient $n$ as usual when
sampling higher frequencies.  

\begin{figure}
  \begin{center}
    \psfig{file=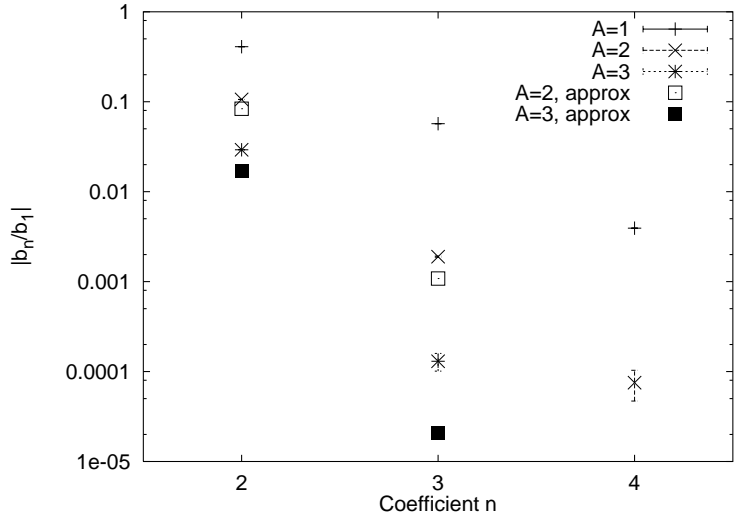,width=4in}
  \end{center}
  \caption{The ratios of Fourier coefficients $\lvert b_n/b_1\rvert$
    plotted against $n$ for the three smallest loop areas.  The
    squares show the same values in the uncorrelated loop approximation.}
  \label{fig:fc_against_n}
\end{figure}

In fig.~\ref{fig:fc_against_n} we show $\lvert b_n/b_1\rvert$ for the
three smallest areas (given in terms of the lattice spacing $a$)
plotted against $n$ on a logarithmic scale: $\lvert b_4/b_1\rvert$ at
area three is omitted because of large errors.  The data is from a
$12^4$ lattice at $\beta=2.5$.  Plotted in this fashion, the data
shows two things.  First, $b_n$ drops off faster for larger $n$.
Second, the rate at which this happens increases for larger loops, as
witnessed by the increased slope.  Even by area $3a^2$ the value for
$n=3$ is down by a factor over a hundred on $n=2$.

Thus the $b_1$ term is the only one contributing to the centre
dominance result, eq.~(\ref{eq:cdexpn}), for loops of even moderate
size and hence the result is the same, up to a trivial normalisation
factor, as that from the full Wilson loop for all medium and large
loops.

In fig.~\ref{fig:fc_against_n}, we have also shown the ratios for
$n=2$ and 3 using the uncorrelated loop approximation, for comparison.
As we have used the value in the plaquette distribution as the
starting point for the higher coefficients, the points for area $1a^2$
are chosen to be the same as in the full case.  However, the trend
thereafter is similar, too.  It has come from two places: first, the
intrinsic behaviour of the model seen above, where higher $b_n$ are
more highly suppressed in larger loops; secondly, the fact that the
initial coefficients $b_n$ for higher $n$ are already smaller and so
powers of higher $b_n$ disappear faster: only the first of the two is
an explicit prediction of the model.

\subsection{Behaviour in higher representations}

It turns out we are able to explain the results in the previous
section as the effect of higher representation loops.  As an
alternative to eqn.~(\ref{eq:cdexpn}), where we expanded the
expectation value in the centre dominance picture in terms of Fourier
coefficients, we could equally well have performed an expansion in
terms of the characters of various representations; our conclusion
would then have been that only the lowest half-odd-integer
representation contributed for large loops.  In fact, 
the expansions are the same up to constant factors.

To show this, we shall define the character of the
representations in terms of the corresponding traces of the Wilson
loop as
\begin{equation}
  \chi_m(W_0) \equiv (m+1)\Tr_{\frac{m}{2}}W,
\end{equation}
where the representation has isospin $m/2$, so that the $\Tr$ operator
is defined to contain the normalisation factor usually used in Monte
Carlo calculations.  The standard recurrence relation between the
characters for SU(2) is
\begin{equation}
  \chi_{m+1}(W_0) = \chi_m(W_0)\chi_1(W_0) - \chi_{m-1}(W_0).
  \label{eq:charrec}
\end{equation}

We shall now show that using our variable $\phi$, defined in
terms of the fundamental representation as
$\Tr_{1/2}W\equiv W_0 \equiv \Sphi$,
the characters for odd and even $m$ can be written,
\begin{equation}
  \begin{aligned}
    \chi_{2(n-1)}(W_0) &\equiv
    (-1)^{n-1}\frac{\cos(n-\frac{1}{2})\phi}{\Cphi}, \\
    \chi_{2n-1}(W_0) &\equiv (-1)^{n-1}\frac{\sin n\phi}{\Cphi}.
  \end{aligned}
  \label{eq:charphi}
\end{equation}
It is to be assumed that the value of the character at
$\phi=\pm\pi$ is found by taking the limit, although we have already
noted that the distribution must vanish there from group-theoretic
considerations.  Note that $\chi_0(W_0) \equiv 1$ and $\chi_1(W_0)
\equiv 2\Sphi$ as expected.  Now we must prove that the following
versions of eqn.~(\ref{eq:charrec})
hold,
\begin{equation}
  \begin{aligned}
    \chi_{2n}(W_0) + \chi_{2n-2}(W_0) &= \chi_{2n-1}(W_0)\chi_1(W_0), \\
    \chi_{2n+1}(W_0) + \chi_{2n-1}(W_0) &= \chi_{2n}(W_0)\chi_1(W_0).
  \end{aligned}
\end{equation}
This can easily be done by insertion of the formulae~(\ref{eq:charphi})
into the respective left hand sides, whence standard trigonometric
relations give the required results.  Hence by induction the
equations~(\ref{eq:charphi}) are true for all $n\ge1$.

Now we work out the expectation values of the traces using the
distribution.  For the representations with $j=n-1/2$, $n\ge1$,
\begin{equation}
  \langle\Tr_{n-\half}W\rangle = \frac{1}{2n}
  \int_{-\pi}^{\pi} \rho(W_0) \chi_{2n-1}(W_0)\Cphi\frac{d\phi}{2}
  = \frac{(-1)^{n-1}\pi b_n}{4n},
  \label{eq:repexpect}
\end{equation}
after inserting the Fourier expansion of eqn.~(\ref{eq:fourier}) and
the second of eqns.~(\ref{eq:charphi}).
Likewise for the representations $j=n-1$, $n\ge1$, we have,
\begin{equation}
  \langle\Tr_{n-1}W\rangle = \frac{(-1)^{n-1}\pi a_n}{4(n-1/2)},
\end{equation}
where $n=1$ and 2 correspond to the identity and
adjoint representations respectively.  Hence, as claimed, there is a
one-to-one correspondence between the Fourier terms and the
representations of the gauge group.  (The basic point about the
relationship between the sign of the loop and the character expansion
was made in ref.~\cite{AmGr98}.)

This means that the behaviour of the coefficients seen in
fig.~\ref{fig:fc_against_n} is entirely consistent with the Casimir
scaling hypothesis, recently discussed in the context of centre
dominance~\cite{FaGr97}, which notes that higher representations
$j=m+1/2$ have a string tension roughly proportional to the quadratic
Casimir operator,
\begin{equation}
  K_j \approx K_C \times j(j+1).
  \label{eq:casscale}
\end{equation}
A larger string tension causes a stronger area law fall off, and hence
from eqn.~(\ref{eq:repexpect}) the Fourier coefficients $b_m$ decay
faster in the loop area.  Combined with eqn.~(\ref{eq:cdexpn}), it can
be seen that Casimir scaling implies centre
dominance in the $j=1/2$ Wilson loops.  This result would be rigorous
if within Casimir scaling one included the assumption that all
loops, not just those in the region of linear confinement, were
suppressed by a similar factor.  As the ratio of string
tensions between $j=3/2$ and $j=1/2$ is already 5 in
eqn.~(\ref{eq:casscale}), it is not surprising that centre dominance
holds so well.

Combining these results with the uncorrelated loop calculation in
eqn.~(\ref{eq:fmult}), we see that in the strong coupling limit all
half-odd-integer representations have an area law behaviour similar to
eqn.~(\ref{eq:arealaw}),
\begin{equation}
\langle \Tr_{n-\half}W(A)\rangle = \left(\frac{(-1)^{n-1}\pi
    b_n}{4n}\right)^A.
\label{eq:halfreparea}
\end{equation}
Indeed, we have also an area law in the integer representations,
\begin{equation}
  \langle \Tr_{n-1}W(A)\rangle = \left(\frac{(-1)^{n-1}\pi
      a_n}{4(n-\half)}\right)^A.
  \label{eq:wholereparea}
\end{equation}
This is perhaps more of a surprise, as one often encounters the
statement `strong coupling predicts a perimeter law behaviour in the
adjoint representation'.  The point is simply that the random
plaquette model we have used sees only the planar contribution to the
adjoint Wilson loop~\footnote{I am grateful to Jeff Greensite for
  making this point clear to me.}.  The perimeter law term comes from
a closed tube of plaquettes surrounding the loop, to which the planar
plaquettes do not contribute in this approximation.  In real loops one
expects both to be present, hence at large $R$ and $T$ the perimeter
law will eventually dominate.  We shall later return to the question
of centre dominance in adjoint loops.

\subsection{Centre projection and positive plaquettes}

In the last few years, detailed investigations of the importance of
Z(2) degrees of freedom have been made by projecting the full SU(2)
links to Z(2).  One then makes plaquettes from these projected links,
creating so-called projection vortices~\cite{DDFa97a,DDFa97b}; a
planar Wilson loop is simply a product of these over the minimal area.

As we have shown (see for example eqn.~(\ref{eq:w0b1})), any
observable contributing to the Wilson loop expectation value must come
from the odd part of the distribution, so that it changes sign when
the relevant loop does.  It is natural to suppose the projection
vortices are related, by some effect of the projection,
to signs of Wilson loops, since that is where the effect of the
centre appears physically.  If this is the case, then the observation
that no vortices means no area law is trivial: as we saw in discussing
the loop distributions, no odd-sign behaviour means, in analytical
terms, no contribution to the real expectation value.  This does not
mean that the relationship between the vortices and
confining behaviour is trivial --- the claim is that the
physical string tension can be seen at particularly short distances in
centre projection --- merely that the presence of both positive and
negative signs is necessary to observe the dynamics.

To make this a little clearer, consider the graphs showing the
expectation value of Wilson loops separated according to whether they
contain odd or even numbers of projection vortices in, for example,
fig.~8 of ref.~\cite{DDFa97a}: the Wilson loops separated in this way
show no confining behaviour.  Indeed, the fact that the Wilson loops
go to zero at all is because the projection vortices do not exactly
correspond to the signs of the full loops: suppose we look instead at
the latter.  As we have already seen, the distribution approaches
(\ref{eq:limiting}) for large loops, and if we consider the part of
the distribution with a positive value of the Wilson loop, we can see
that large Wilson loops approach a constant value,
\begin{equation}
  \langle \lvert W_{0l}\rvert\rangle = \int_{-1}^1 \lvert W_0\rvert
  \rho_l(W_0)dW_0
  =\frac{2}{\pi} \int_{-1}^1 \lvert W_0\rvert(1-W_0^2)^{1/2}dW_0
  =\frac{4}{3\pi},
  \label{eq:evlimit}
\end{equation}
and the negative loops taken alone correspondingly approach the limit
$-4/(3\pi)$.  This is a property of the even part of the distribution,
unrelated to the odd part where $\langle W_0\rangle$ is to be found.
(However, the adjoint string tension lives in the even part of the
distribution: see below.)

The relationship between thick vortices and the sign of the loop is
clearer in the positive plaquette model~\cite{MaPi82}.  Here,
plaquettes are constrained to have a positive value: this can be
implemented by a simple accept/reject test in Monte Carlo updating.
This model appears to have all the right properties to be an
alternative lattice regularisation of SU(2)~\cite{FiHe94}.  There
are no thin vortices, which are simply lines of negative
plaquettes, so that through the commutativity property of the
centre any Wilson loop with a positive (negative) sign must contain
an even (odd) number of thick vortices.

\begin{figure}
  \begin{center}
    \psfig{file=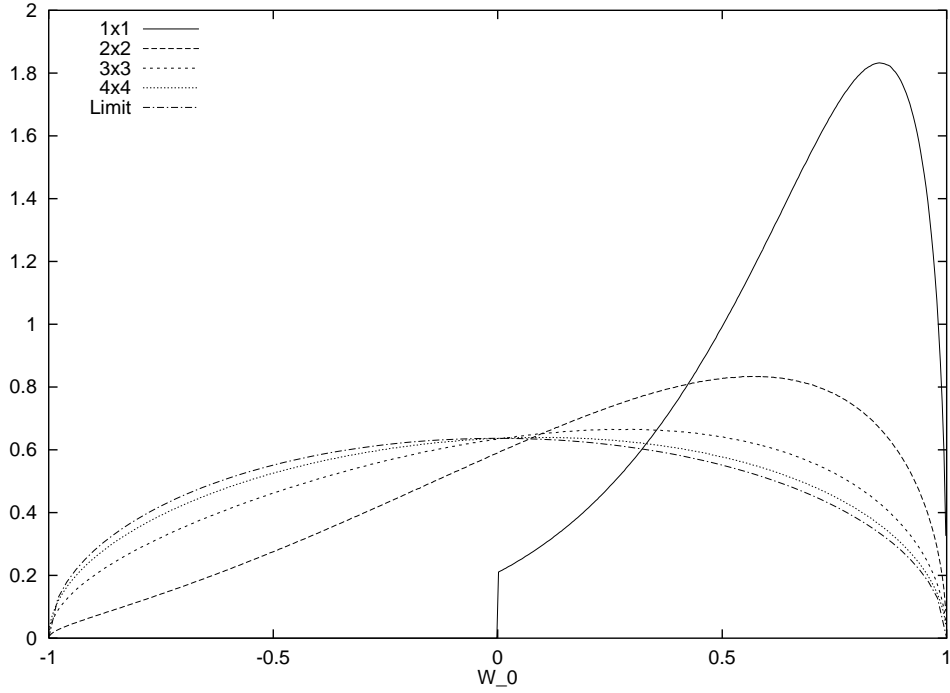,width=3.7in,angle=270}
  \end{center}
  \caption{Monte Carlo distribution of Wilson loops on a $12^4$
    lattice at $\beta=1.9$ for the positive plaquette model,
    corresponding otherwise to figure~\protect\ref{fig:wl_dist}.}
  \label{fig:wl_ppm_dist}
\end{figure}

We have investigated the Wilson loop distribution in this theory, and
show the results in fig.~\ref{fig:wl_ppm_dist}.  The coupling here is
$\beta=1.9$, found to be in the scaling region for this
theory~\cite{FiHe94}.  The approach to the large loop region is,
unsurprisingly, very similar to that shown in standard SU(2) in
fig.~\ref{fig:wl_dist}.

The sign of the Wilson loop here counts thick vortices exactly and so
the expectation values of Wilson loops with only even or odd numbers
of thick vortices must tend to $\pm4/3\pi$, irrespective of
whether the vortices are responsible for the dynamics, i.e.\ %
irrespective of centre dominance.  Our point here is simply that one
\textit{must} consider loops with odd and even numbers of vortices
together, as separately they can have no interpretation in terms of
$\langle W_0\rangle$.

\begin{figure}
  \begin{center}
    \psfig{file=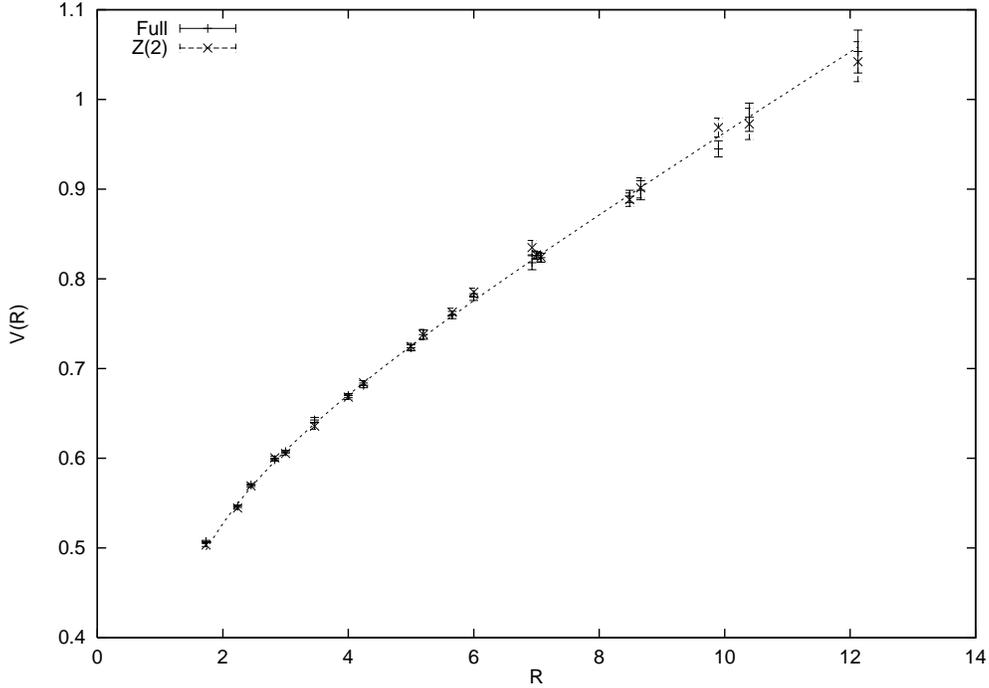,width=3.7in,angle=270}
  \end{center}
  \caption{The heavy quark-antiquark potential from the positive
    plaquette model at $\beta=1.8$, in lattice units.  The dashed line
    is a three-parameter fit.}
  \label{fig:ppm_pot}
\end{figure}

We can also use the positive plaquette model to examine the question
of whether the results from the Z(2) part agree with the full results
when only thick vortices are present, or whether, on the other hand,
both thick and thin vortices are required to produce the full result.
Fig.~\ref{fig:ppm_pot} shows the potential between a heavy quark and
an antiquark calculated from 150 configurations on a $16^4$ lattice at
$\beta=1.8$ for both the full results and those using only the signs
of the Wilson loops.  This figure corresponds to figures 7 and 8 of
ref.~\cite{KoTo97} which show results in standard pure SU(2); our
method is the same, i.e.\ $V(R)$ is calculated from
\begin{equation}
V(R) = \lim_{T\to\infty} -\log\frac{W(R,T+1)}{W(R,T)}
\label{eq:potdef}
\end{equation}
with the signal improved by using fuzzing on spatial links only ---
this does not affect the presence of vortices in the $(R,T)$ plane.
To enable a comparison with the standard case we have also made a fit
to the form
\begin{equation}
  V(R) = V_0 + KR - e/R;
\end{equation}
no correction for lattice artifacts has been made and 
the single correlated fit included all data with $T \ge 3$ and
$R \ge 2.5$ for the full loops only (i.e.\ not using the Z(2)
projected values). We obtained $V_0=0.572(6)$, $K=0.0410(9)$ and $e=
0.261(11)$, corresponding to $\beta$ slightly below 2.5 for standard
SU(2).

It is clear from fig.~\ref{fig:ppm_pot} that, as with standard SU(2)
in ref.~\cite{KoTo97}, the Z(2) part carries the physics; a similar
picture (not shown) holds for $\beta=1.9$ where the curvature is more
pronounced.  Thus centre dominance is present across the whole size
range, from the Coulomb to the confining region, works even with no
thin vortices present.  This is something of a relief, as the division
into thin and thick vortices is an artifact of the cut-off.  To see
this, consider taking a negative plaquette, i.e.\ part of a thin
vortex, and halving the lattice spacing: on the new lattice, it is no
longer clear whether the loop corresponding to the former plaquette is
(say) part of a thick vortex surrounding four positive plaquettes on
the new lattice, or is negative by virtue of one of the those
plaquettes being negative.  In other words, the new lattice does not
preserve the division into thin and thick vortices of the old one.

\subsection{Confinement from non-interacting vortices}

We will show that a random distribution of non-interacting thick
centre vortices, with no other dynamics, can give an area law in SU(2)
with a string tension of about the right magnitude.  This is the
simplest possible treatment, although it is very much in the spirit of
Nielsen--Olesen flux vortices~\cite{NiOl73,NiOl79}; random vortices
and their confining nature are not new and were considered for example
in~\cite{Ol82}.  However it is now possible to make a quantitative
test of the idea using results from centre vortices.  A similar (but
not identical) calculation appeared in ref.~\cite{FaGr97}; the version
here has the advantage that it makes no reference to the lattice, even
if the underlying assumptions are too simplistic for it to be a true
continuum theory.

We assume that the value of a Wilson loop of area $A$ depends only on
the number of vortices inside, i.e.\ $W_0(A) = W_C(-1)^n$, where $n$
vortices pass through the loop $W_0$ and $W_C$ depends only on the
coupling: this is just the centre dominance proposal, which we have
seen works well in practice. Although the assumption in this form does
not fit in with the Wilson loop distributions we have shown, the
proposal is that the sign alone determines the dynamics, so that the
effect of the rest of the distribution on $\langle W_0\rangle$ can be
absorbed into $W_C$.

We also assume that the vortices are independent from one another, so
that any vortex can pass at a random point through a given area
regardless of the presence of another: this second assumption is
clearly an oversimplification, as there is indeed interaction between
the vortices~\cite{EnLa98}.  We are also assuming the vortices are
infinitely thin, so that there are no edge effects; this, too, is
unphysical, as we have already stressed that the vortices must be
thick to survive in the continuum limit.

Finally, we suppose that the only relevant physical quantity is the
mean density of vortices passing through unit area which we denote by
$p_A$.  In addition, we are using the fact of translational invariance
so that there are equal probabilities for a vortex to pierce any
regions with the same area.  We are also essentially ignoring the fact
that the vortices appear from the gauge dynamics.

\begin{figure}
  \begin{center}
    \psfig{file=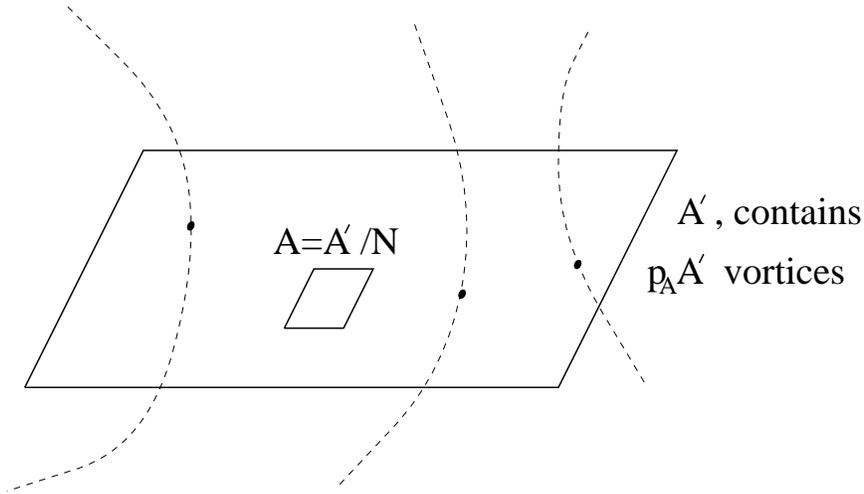,angle=270,width=4.5in}
  \end{center}
  \caption{The random non-interacting vortex model, as described in
    the text.}
  \label{fig:rand_vort}
\end{figure}

Consider an area very much larger than $A$,
$A' = AN$ for $N\gg 1$, which encloses $A$ and through which
exactly $p_AA'$ vortices are assumed to pass, randomly distributed
across the area: a three-dimensional slice is shown in
fig.~\ref{fig:rand_vort}.  For each vortex
independently, there is a probability $1/N$ that it lies inside
$A$. The probability for $n$ of them to lie in the area $A$ is then
given by a binomial distribution,
\begin{equation}
\Pr(\text{$n$ vortices in $A$}) = \binom{p_AA'}{n}\left(\frac{1}{N}\right)^n
\left(1-\frac{1}{N}\right)^{p_AA'-n}.
\end{equation}

The contribution to the Wilson loop is
\begin{equation}
\langle W_0(A) \rangle = W_C \sum_{n=0}^{p_AA'}\binom{p_AA'}{n}(-1)^n
\left(\frac{1}{N}\right)^n\left(1-\frac{1}{N}\right)^{p_AA'-n}
     = W_C \left(1 - \frac{2}{N}\right)^{p_AAN}.
\end{equation}
Letting $N\to\infty$, and noting that the limit as $N$ tends to
infinity of $(1-2/N)^N$ is $e^{-2}$, we have
\begin{equation}
\langle W_0(A)\rangle = W_C e^{-2p_AA},
\end{equation}
which is an area law with string tension $K=2p_A$.  This result was
obtained in a different approximation by ref.~\cite{DDFa98}; in this case
we stress that its validity is based only on the notion of
randomly distributed, non-interacting and (unfortunately) thin
vortices.

A string tension of $K = (440 \text{MeV})^2$ would hence require
$p_A\approx 2.5\>\text{fm}^{-2}$. Determining this quantity in a gauge
invariant fashion is difficult, so we shall use the value calculated
directly from Z(2)-projected vortices in ref.~\cite{LaRe97}: $p_A =
(1.9\pm0.2)\>\text{fm}^{-2}$, close to what we require.  Further, the
same value of the string tension was assumed in that calculation, so
in fact the ratios of the two numbers are independent of the
experimental value of $K$.  Hence the random vortex model predicts a
value of $K$ about three quarters of the measured value (giving for
$\sqrt{K}$ some 85\% of the measured value), the difference
corresponding to three standard deviations.  This is surprisingly good
considering the model in question contained very little mathematics
and did not even take into account the extended nature of the
vortices.  That the centre vortices are physical is supported by the
result of ref.~\cite{LaRe97} that the distribution is renormalisation
group invariant.

\subsection{The adjoint representation}

It has long been argued, and has to some extent entered the folklore,
that the observation of a string tension in the adjoint representation
of SU(2) cannot easily be explained by centre vortices.  This claim is
based on the point that the centre Z(2) of SU(2) is not seen by
adjoint fields.  This is, of course, correct, but this Z(2) is
associated with \textit{thin} centre vortices, which are not expected
to be involved with confinement in the continuum limit --- and as we
saw from the positive plaquette model are not necessary even for
lattice physics.  The surviving symmetry of the adjoint representation
is SU(2)/Z(2), which is just that required for \textit{thick} centre
vortices.  The mathematics of confinement relates to the fact that
with SU(2)/Z(2) one has two types of path in the gauge manifold, one
contractable and one not; the latter are closed due to the
identification of opposite ends of a diameter of the manifold.  This
distinction survives in the adjoint representation.

In the adjoint representation, up to constants irrelevant to the
argument, the action depends on the square of the plaquette value, and
the negative trace of the loops is not seen.  However, the fact that one
cannot measure the signs of the loops in no way blurs the distinction
between the topologically different windings of the field.  One has
simply lost a useful but --- if the theory of thick centre vortices is
correct --- physically unnecessary label, namely the association
between the sign of the loop and the presence of a vortex.

This last remark explains the difficulty with projection vortices,
where factors of $-1$ presumably correspond to a thick centre vortex
with an increasing admixture of thin vortices for stronger coupling.
Such vortices certainly are not present in the adjoint representation,
making investigation of the mechanism considerably more difficult, but
the underlying topology of the gauge manifold which forms the basis
for confinement in the theory is still present.

The point made here is slightly different from the
suggestion~\cite{FaGr97} for reconciling the adjoint string tension
with the vortex picture involving sign flips.  However, the finite
width of the vortices enters in both cases: in that mechanism it was
the origin of string behaviour in higher representations, while here
we have stressed the SU(2)/Z(2) nature of the loops, which must be
spread over a large area so as to be present in the continuum limit.
Thus there may be a connection between the two.

We can show the correlations between the behaviour of adjoint Wilson
loops and the presence of vortices directly.  We have run a simulation
on our $12^4$ lattice at $\beta=2.5$ in which the adjoint loops are
separated according to whether the fundamental loop was positive or
negative, i.e.\ whether there is an even or odd number of vortices
passing through the loop.  This is just the process we criticised
above for the fundamental case; however, here the sign of $W_0$ itself
is not seen by the adjoint trace,
\begin{equation}
  W_0^\mathrm{adj} = (4W_0^2-1)/3,
  \label{eq:adjdef}
\end{equation}
so that any correlation with the
sign involves the SU($N$)/Z($N$) topology associated with the
thick vortices.  Indeed, the adjoint loop probes the even part of the
distribution $\rho(W_0)$ whereas, as we have made clear, 
the expectation values of fundamental loops and their signs only
involve the odd part.

\begin{figure}
  \begin{center}
    \psfig{file=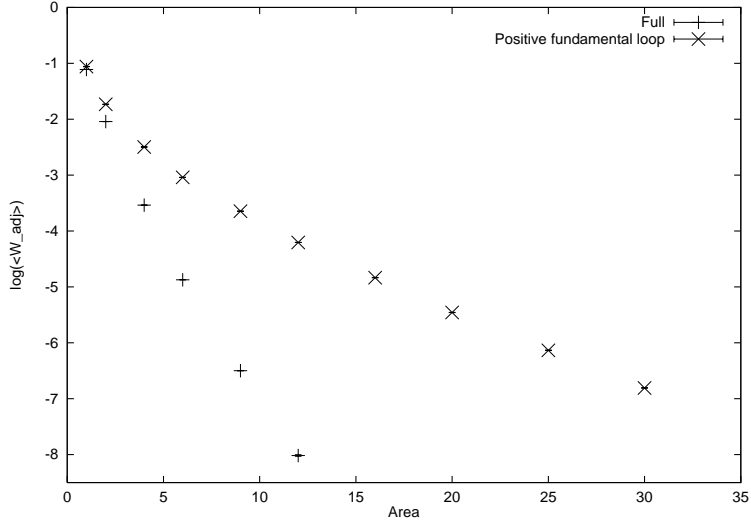,angle=270,width=4in}
  \end{center}
  \caption{Logarithms of adjoint Wilson loops plotted against area.
    Both the complete data, and those loops associated with a positive
    fundamental loop, are shown.}
  \label{fig:log_adj}
\end{figure}

In fig.~\ref{fig:log_adj}, we show the logarithms of the adjoint
Wilson loops, both for the complete data and those with positive
fundamental trace, once more at $\beta=2.5$.  There is clear sign of
an area law behaviour in the former, i.e.\ an adjoint string tension.
The latter have a consistently shallower gradient, in other words the
string tension is significantly lower in those loops with only even
numbers of vortices.  A perfect or near-perfect correlation is not to
be expected since from eqn.~(\ref{eq:adjdef}) that would require all
loops with $W_0 < 0$ also to have $\lvert W_0 \rvert < 1/4$, and the
other way around for positive $W_0$.

Phenomenologically, the source of the correlation is not hard to see
from the fundamental distribution in fig.~\ref{fig:wl_dist}, bearing
in mind that the expectation value we want is just
eqn.~(\ref{eq:adjdef}) integrated over this distribution.  For smaller
loops, the distribution peaks at positive $W_0$ and tails off towards
negative values.  Therefore, large absolute values of $W_0$, and hence
positive contributions to $W_0^\mathrm{adj}$, tend to come
predominantly from positive fundamental loops.  However, the source of
the adjoint string tension itself is rather less easy to fathom by
this method, and for that we appeal to the strong coupling result
of eqn.~(\ref{eq:wholereparea}) with $n=2$.

As this work was being completed, ref.~\cite{FaGr98} appeared, showing
similar correlations between the presence of vortices and the adjoint
string tension.

\section{Summary}

The major results of this paper are as follows:
\begin{itemize}
 \item We considered the distribution $\rho(W_0)dW_0$ of Wilson loops
  in SU(2); for loops larger than the correlation length of the field
  $\rho(W_0)$ reaches the limiting form $(2/\pi)\sqrt{1-W_0^2}$.

 \item We applied the Fourier decomposition in eqn.~(\ref{eq:fourier});
  the expectation value of the loop is $\langle W_0\rangle=\pi
  b_1/4$, in other words it only probes the first odd coefficient of
  the distribution in this parametrisation; the even portions of
  $\rho(W_0)$ do not contribute.
  
 \item In the simple approximation where all Wilson loops are
  uncorrelated, and we use the measured plaquette distribution as
  input, the coefficient $b_1$ and hence $\langle W_0\rangle$ obey the
  strong coupling behaviour, eqn.~(\ref{eq:arealaw}).  Larger
  coefficients $b_n$ are suppressed by an
  additional factor $1/n$ for each multiplication by a Wilson loop of
  the same size.  This approximation also shows an area law behaviour
  for the adjoint representation.

 \item The Z(2) expectation value contains all $b_n$, not just $b_1$.
  We found that for $n>2$ the coefficients were strongly suppressed
  in Monte Carlo results, showing the validity of centre dominance.

 \item The Fourier terms correspond, term by term, with
  the expectation value of the Wilson loop in the gauge group
  representations.  The faster decay of higher representation loops
  therefore gives an explanation for centre dominance.

 \item We stressed that care needs to be taken when looking at results
  in Z(2)-projection (whether the links are projected or we simply
  look at the sign of the loops): the expectation values of the
  negative and positive parts separately are trivially seen to be
  non-confining, from the basic properties of the distribution.

 \item We showed by considering the positive plaquette model that the
  absence of thin vortices, which are chains of negative plaquettes,
  does not change the centre dominance behaviour.

 \item We showed that even a simple physical model of a random gas of
  centre vortices with no interactions gave three quarters of the
  observed string tension from the measured vortex density.

 \item We pointed out that there is no essential difficulty in having
  \textit{thick} centre vortices explain an adjoint string tension, as
  both the thick vortices and the adjoint representation are
  associated with an SU($N$)/Z($N$) symmetry --- the theory of centre
  dominance does not directly involve the centre Z($N$), as has
  sometimes been incorrectly assumed.

 \item We showed that there is a correlation between the adjoint
  string tension and the sign of the corresponding fundamental Wilson
  loop, even though the latter does not directly affect the adjoint
  loop: this makes it plausible that thick vortices are involved here,
  too.  This correlation is easily seen by looking at the Wilson loop
  distributions.
\end{itemize}

These results seem to show that there is indeed physics contained in
the centre dominance picture.  Of course, the results for other
pictures of confinement need to be borne in mind.  For example, a lot
of physics has also emerged from the maximally Abelian monopole
condensation (Abelian dominance, or dual superconductor vacuum)
picture~\cite{abmono}; see e.g.\ the proceedings of Lattice
'97~\cite{Latt97} for recent progress.  A way in which Abelian
monopoles might be related to centre vortices has been
proposed~\cite{DDFa97b}.  It is interesting to note that both models
are based on topological mechanisms using homotopy classes, those of
SU(2)/Z(2) or U(1), not manifest in the full gauge group, and one can
speculate that there is a deeper connection.  There are of course
other models, too, such as the anti-ferromagnetic vacuum~\cite{Po96}.
The arguments for the fundamental importance of Z(N) in SU(N) given by
't~Hooft twenty years ago~\cite{Ho78} are still valid and should be
born in mind, but the debate looks set to continue.

For further progress, one needs a better way of asking which
properties are the most fundamental.  Work is currently in progress by
another group to try to understand the correlations between the
monopole picture and instanton effects~\cite{IlMa98}; centre dominance
is presumably another ingredient which needs to be taken into account.
How far our results here really reflect dynamics involving the centre
of the gauge group can presumably be clarified by looking at SU(3).
Recent results~\cite{KoTo98} suggest the corresponding behaviour of
Wilson loops for the centre Z(3) is indeed found in the SU(3) case.


\begin{thebibliography}{99}
\bibitem{Co79} J.M.~Cornwall, \np{B157}{1979}{442}.
\bibitem{Ho79} G.~'t Hooft, \np{B153}{1979}{141}.
\bibitem{MaPe79} G.~Mack and V.B.~Petkova, \ap{123}{1979}{442};
 (1979); \ib{125}{1980}{117}; \zp{C12}{1982}{177}.
\bibitem{DDFa97a} L.~Del Debbio, M.~Faber, J.~Greensite and \v
 S. Olejn\'\i k, \prev{D55}{1997}{2298}, hep-lat/9610005.
\bibitem{DDFa97b} L.~Del Debbio, M.~Faber, J.~Greensite and \v
 S. Olejn\'\i k, talk at the NATO Workshop ``New Developments in
 Quantum Field Theory'', June 1997, Zakopane, Poland, hep-lat/9708023.
\bibitem{DDFa98} L.~Del Debbio, M.~Faber, J.~Giedt, J.~Greensite and
\v S. Olejn\'\i k, hep-lat/9801027.
\bibitem{KoTo97} Tam\'as G. Kov\'acs and E.T.~Tomboulis,
 \prev{D57}{1998}{4054}, hep-lat/9711009; Nucl.\ Phys.\ B (Proc.\ Suppl.)
 63 (1998) 534, hep-lat/9709042.
\bibitem{BaSc95} G.~Bali, C.~Schlichter and K.~Schilling,
 \prev{D51}{1995}{5165}, hep-lat/9409005.
\bibitem{MaPo82} Yu.M.~Makeenko, M.I.~Polikarpov, A.I.~Veselov,
 \pl{B118}{1982}{133}.
\bibitem{BeMa84} T.I.~Belova, Yu.M.~Makeenko, M.I.~Polikarpov,
 A.I~Veselov, \np{B230}{1984}{473}.
\bibitem{AmGr98} J.~Ambjorn, J.~Greensite, J.~High Energy Phys.\ 05:004,
 1998, hep-lat/9804022.
\bibitem{MaPi82} G.~Mack and E.~Pietarinen, \np{B205 [FS5]}{1982}{141}.
\bibitem{FiHe94} J.~Fingberg, U.M.~Heller and V.~Mitryushkin,
 \np{B435}{1995}{311}, hep-lat/9407011.
\bibitem{NiOl73} H.~Nielsen and P.~Olesen, \np{B61}{1973}{45}
\bibitem{NiOl79} H.~Nielsen and P.~Olesen, \np{B160}{1979}{380}.
\bibitem{Ol82} P.~Olesen, \np{B200 [FS4]}{1982}{381}.
\bibitem{FaGr97} M.~Faber, J.~Greensite and S.~Olejn\'\i k,
 \prev{D57}{1998}{2603}, hep-lat/9710039.
\bibitem{LaRe97} Kurt Langfeld, Hugo Reinhardt and Oliver Tennert,
 \pl{B419}{1998}{317}, hep-lat/9710068.
\bibitem{EnLa98} M.~Engelhardt, K.~Langfeld, H.~Reinhardt and
 O.~Tennert, hep-lat/9801030.
\bibitem{FaGr98} M.~Faber, J.~Greensite and \u S. Olejn\'\i k,
 hep-lat/9807008.
\bibitem{abmono} G.~'t Hooft, in EPS Conference on High Energy Physics,
   Palermo 1975, A. Zichichi, ed.; S.~Mandelstam,
   \prep{23C}{1976}{245}; G.~Parisi, \prev{D11}{1975}{971}.
\bibitem{Latt97} Topology and Confinement parallel session,
 proceedings of Lattice '97, C.T.H~Davies, I.M.~Barbour, K.C.~Bowler,
 R.D.~Kenway, B.J.~Pendleton and D.G.~Richards, eds., Nucl.\ Phys.\ B
 (Proc Suppl.) 63 (1998) 465.
\bibitem{Po96} J.~Polonyi, talk at the Workshop on Quark Confinement
 and Hadron Spectrum II, Como, Italy, June 1996, hep-lat/9610030.
\bibitem{Ho78} G.~'t~Hooft, \np{B138}{1978}{1}.
\bibitem{IlMa98} E.-M.~Ilgenfritz, H.~Markum, M.~M\"uller-Preussker,
 W.~Sakuler and S.~Thurner, hep-lat/9804031.
\bibitem{KoTo98} Tam\'as G. Kov\'acs and E.T.~Tomboulis,
 hep-lat/9807027.
\end{thebibliography}
\end{document}